\documentclass[12pt]{article}
\usepackage{amsmath, amsthm, amssymb}
\usepackage[body={0in, 0in}, bottom=1in, top=1in,left=1in,right=1in]{geometry}
\usepackage{graphicx}
\usepackage{hyperref}
\usepackage{threeparttable}

\begin{document}

\title{A Survival Copula Mixture Model for Comparing Two Genomic Rank Lists}
\markboth%
{YY Wei and HK Ji}
{Genomic Rank Lists Comparison}

\author{Yingying Wei, Hongkai Ji\\ Johns Hopkins University Bloomberg School of Public Health}
\date{}

\maketitle

\begin{abstract}
Analyses of high-throughput genomic data often lead to ranked lists of genomic loci. How to characterize concordant signals between two rank lists is a common problem with many applications. One example is measuring the reproducibility between two replicate experiments. Another is to characterize the interaction and co-binding between two transcription factors (TF) based on the overlap between their binding sites. As an exploratory tool, the simple Venn diagram approach can be used to show the common loci between two lists. However, this approach does not account for changes in overlap with decreasing ranks, which may contain useful information for studying similarities or dissimilarities of the two lists. The recently proposed irreproducible discovery rate (IDR) approach compares two rank lists using a copula mixture model. This model considers the rank correlation between two lists. However, it only analyzes the genomic loci that appear in both lists, thereby only measuring signal concordance in the overlapping set of the two lists. When two lists have little overlap but loci in their overlapping set have high concordance in terms of rank, the original IDR approach may misleadingly claim that the two rank lists are highly reproducible when they are indeed not. In this article, we propose to address the various issues above by translating the problem into a bivariate survival problem. A survival copula mixture model is developed to characterize concordant signals in two rank lists. The effectiveness of this approach is demonstrated using both simulations and real data.
\end{abstract}


\section{Introduction}
\label{s:intro}

Analyses of high-throughput genomic data often produce ranked lists of genomic loci. Two examples are lists of differentially expressed genes from an RNA-seq or microarray experiments \cite{Reflimma} and lists of transcription factor (TF) binding peaks from ChIP-seq data \cite{RefCisgenome}. In each list, loci are ranked based on scores such as p-values, false discovery rates (FDR) \cite{RefFDR,rStorey1,rStorey2} or other summary statistics. When two such lists are available, a common problem is to characterize the degree of concordance between them. Below are two examples.

\begin{itemize}
\item \textit{Characterizing co-binding of two transcription factors}: ChIP-seq data are collected for two different TFs. For each TF, an initial data analysis yields a list of peaks along the genome representing its putative binding regions. In order to characterize whether the two TFs collaborate and how they interact with each other, one wants to compare the two peak lists to answer the following questions: (1) What proportion of the true binding sites are shared by the two TFs? (2) How does this proportion change as one moves from high quality peaks to low quality ones?

\item \textit{Assessing reproducibility of scientific findings}: Gene expression data for the same biological system are collected independently by two different laboratories. Each lab collects the data using its own platform and protocol. The data from each lab contain gene expression profiles for two biological conditions, each with multiple replicate samples. Each lab analyzes its own data to generate a list of differentially expressed genes. One wants to compare the differential gene lists from the two labs to determine which differential genes are likely to be reproducible by other labs.
\end{itemize}

In both these scenarios, perhaps the best way to compare two datasets is to model it at the raw data level. Whenever possible, directly comparing or modeling the raw data may allow one to keep most of the information. However, this is not always easy or feasible. For instance, sometimes genomic rank lists are published without releasing the raw data to protect confidentiality of research subjects. Sometimes, one may want to compare his/her own data with thousands of other datasets in public repositories such as ENCODE \cite{RefEncode}, modENCODE \cite{RefmodENCODE}, and Gene Expression Omnibus (GEO) \cite{RefGEO}. Analyzing all the raw data in these databases is a huge undertaking that requires significant amount of resources. This is often beyond the capacity of an individual investigator, and it may not be justified based on the return. In those situations, comparing two datasets based on the readily available rank lists may be preferred. Sometimes, this may be the only solution. This article considers analysis issues in this scenario.

As an exploratory tool, the simple Venn diagram approach is widely used to show the overlap between two genomic loci lists. However, this approach does not consider the concordance or correlation of ranks between the two lists. A feature commonly seen in genomic rank lists is that the top ranked loci are more likely to be true signals. Signals are more likely to be reproduced in independent studies than noise; therefore, they tend to be correlated between different datasets.
Because of this, the concordance between the two rank list is a function that changes with the rank of the loci. This information is not reflected in a Venn diagram.
To address this limitation, Li et al. recently proposed a method to measure the concordance of two rank lists as a function of rank. They developed a
Gaussian copula mixture model to assign a reproducibility index, irreproducible discovery rate (IDR), to each locus. The IDR analysis produces a concordance curve rather than a scalar number to measure the overlap between two lists. This approach is semiparametric and invariant to monotone transformations of the scores used for ranking \cite{RefLi}. In principle, IDR is a model based version for one minus the correspondence at the top (CAT) plot proposed by Irizarry et al.\cite{RefCAT}. The original authors of IDR demonstrated their method using an application where they evaluated the reproducibility of different ChIP-seq peak callers by comparing the peak calling results from two replicate experiments.

Although the IDR approach represents a significant advance compared to the simple Venn diagram analysis, it also has limitations. Importantly, the Gaussian copula mixture model in the original IDR approach requires one to know the ranks of each locus in both lists. However, many loci occur only in one list.
As a result, to perform the IDR analysis, Li et al. first filtered out all loci that were reported in only one rank list. Loci are included in the IDR analysis only if they appear in both lists. As Li et al. reported, for the real data they analyzed (which are peak lists from replicate ChIP-seq experiments), only ``23-78\% of peaks are retained for this analysis'' \cite{RefLi}. As such, the original IDR analysis only characterizes the signal concordance for a subset of loci that are reported in both lists. Attempting to interpret the resulting IDR as a reproducibility measure for the whole dataset could be misleading. It is possible that the two original loci lists have little overlap and, therefore, low reproducibility, but the loci in their overlapping set (i.e., the loci shared by both lists) are highly correlated in terms of their relative ranking. In such a situation, the IDR computed using only the overlapping loci may misleadingly suggest high reproducibility of the two datasets. This is a limitation caused by ignoring list-specific loci, and can only be addressed by bringing them back into the analysis.

Here we propose a Survival COPula mixture model, SCOP, to tackle the general problem of comparing two genomic rank lists.
This new approach allows one to include the list-specific loci in the analysis when evaluating the signal concordance between the two datasets.
For loci that occur only in one list, we treat the scores used for ranking (e.g., p-values or FDRs) in the other list as censored data.
In this way, we translate the problem into a bivariate survival problem. Although many works have been done in the area of estimating correlation structure of bivariate failure times in survival analysis \cite{RefFrailty,rFan,RefMultSurv,RefHougaard,RefLiYi,RefNan,RefCopula,RefOakes,rLouis}, none of them considered the issue specific to genomic data. In genomic applications, the higher ranked loci are more likely to be true signals. Thus, adopting the traditional survival analysis terminology, earlier failure time is of higher interest. Built upon Li et al, our survival copula mixture model attempts to borrow strength from both the copula mixture model and the survival analysis. The benefit is that it can better characterize the overlap and concordance between two rank lists.

The article is organized as follows. In Section 2, we introduce the survival copula mixture model and discuss its connection to survival analysis. Section 3 uses simulations to demonstrate our method and compare it with the other alternatives. We apply our method to two real ChIP-seq datasets example in Section 4. We then conclude the article with discussions in Section 5.

\section{Method}
\label{s:model}

\subsection{Data structure}
Consider two genomic loci lists such as lists of differentially expressed genes from two RNA-seq experiments or lists of transcription factor binding regions from two ChIP-seq experiments. In each list $j$ ($\in \left\{1,2\right\}$), loci are rank ordered based on a score such as a p-value or an FDR. Let $T_{i,j}$ be the score for locus $i$ in list $j$. Without loss of generality, we assume that smaller score (e.g., smaller FDR) represents a higher significance. Often, a locus is reported in list $j$ only when its score passes a cutoff $C_{j}$. Thus, all loci in list 1 satisfy $T_{i,1} \leq C_{1}$, and any locus with $T_{i,1} > C_{1}$ is not reported. Similarly, list 2 contains loci for which $T_{i,2} \leq C_{2}$. A locus may be reported in both lists, in one list only, or in none of the lists. Each list may contain a certain amount of noise or false positives in addition to signals. By comparing the two lists, the goal is to characterize the degree of concordance of the signals from the two datasets, and how the concordance varies as one moves from the top ranked loci to those lower ranked.

To analyze these data, we borrow the idea of IDR. However, instead of excluding loci that occur in only one list from the analysis, we retain all loci that occur in any of the two lists. If a locus does not appear in one list, its score in that list is labeled as missing. This creates missing data, but the data here are not missing completely at random. For example, if rank list 1 uses an FDR cutoff of 0.1, then we know that for any loci in list 2 but not in list 1, their missing FDR in list 1 are indeed greater than 0.1. In other words, the data we observe are right truncated. This naturally translates the problem into a survival problem with right censoring data.

Figure 1(a) shows a numerical example. The figure displays two ChIP-seq peak lists ranked according to FDR. Region 2 passes the FDR cutoff for both lists, but region 1 only appears in the peak list for TF A. It is absent in the peak list for TF B since its FDR in that dataset is higher than the 0.1 cutoff. Rather than excluding region 1 from the analysis, we retain it and encode the data using ``observed survival time'' and ``censoring indicator'' adopting the terminology in survival analysis. The ``observed survival time'' is defined as $X_{i,j}=min\{T_{i,j},C_{j}\}$, and the ``censoring indicator'' is defined as $\delta_{i,j}=I(T_{i,j}\leq C_{j})$. In this example, the observed survival time for region 1 in peak list B is 0.1, and the censoring indicator is equal to zero indicating that the data is censored.
Intuitively, the original IDR approach by \cite{RefLi} only models the red points (i.e., cases with complete data) in Figure 1(b), whereas our new approach attempts to use information from all data points regions II, III, and IV. Later we will show that compared to the original IDR calculation which excludes the list-specific loci, including them as censored data in our model will provide more information.

\subsection{The SCOP Model}
Let $f_j(t)$ be the probability density function for $T_{i,j}$. $S_j(t)=P(T_{i,j}\geq t)$ is the corresponding survival function. For any bivariate random variables, there exists a copula, which is invariant under monotone transformation for the marginal distribution \cite{RefCopula}. Based on this, we use two latent random variables $Z_{i,1}$ and $Z_{i,2}$ to characterize the relationship between $T_{i,1}$ and $T_{i,2}$. For each $j$, $Z_{i,j}$ is assumed to follow a Gaussian mixture distribution. $G_j(z)=P(Z_{i,j}\geq z)$ represents the survival function for the latent variable $Z_{i,j}$. The latent variables ($Z_{i,1}$ and $Z_{i,2}$) and the observed scores ($T_{i,1}$ and $T_{i,2}$) are linked through a monotone transformation $S_j(t_{i,j})=G_j(z_{i,j})$.
Let $g_1(z)$ denote the density function for $Z_{i,1}$. It is assumed that this density is a mixture of a noise component $g_{10}\sim N(0,1)$ and a signal component $g_{11}\sim N(\mu_1,\sigma_1^2)$, where $\mu_1<0$. Similarly, the density function for $Z_{i,2}$, $g_2(t)$, is assumed to be a mixture of noise $g_{20}\sim N(0,1)$ and signal $g_{21}\sim N(\mu_2,\sigma_2^2)$ where $\mu_2<0$.

\begin{figure}
\begin{center}
\includegraphics[width=1.1\textwidth]{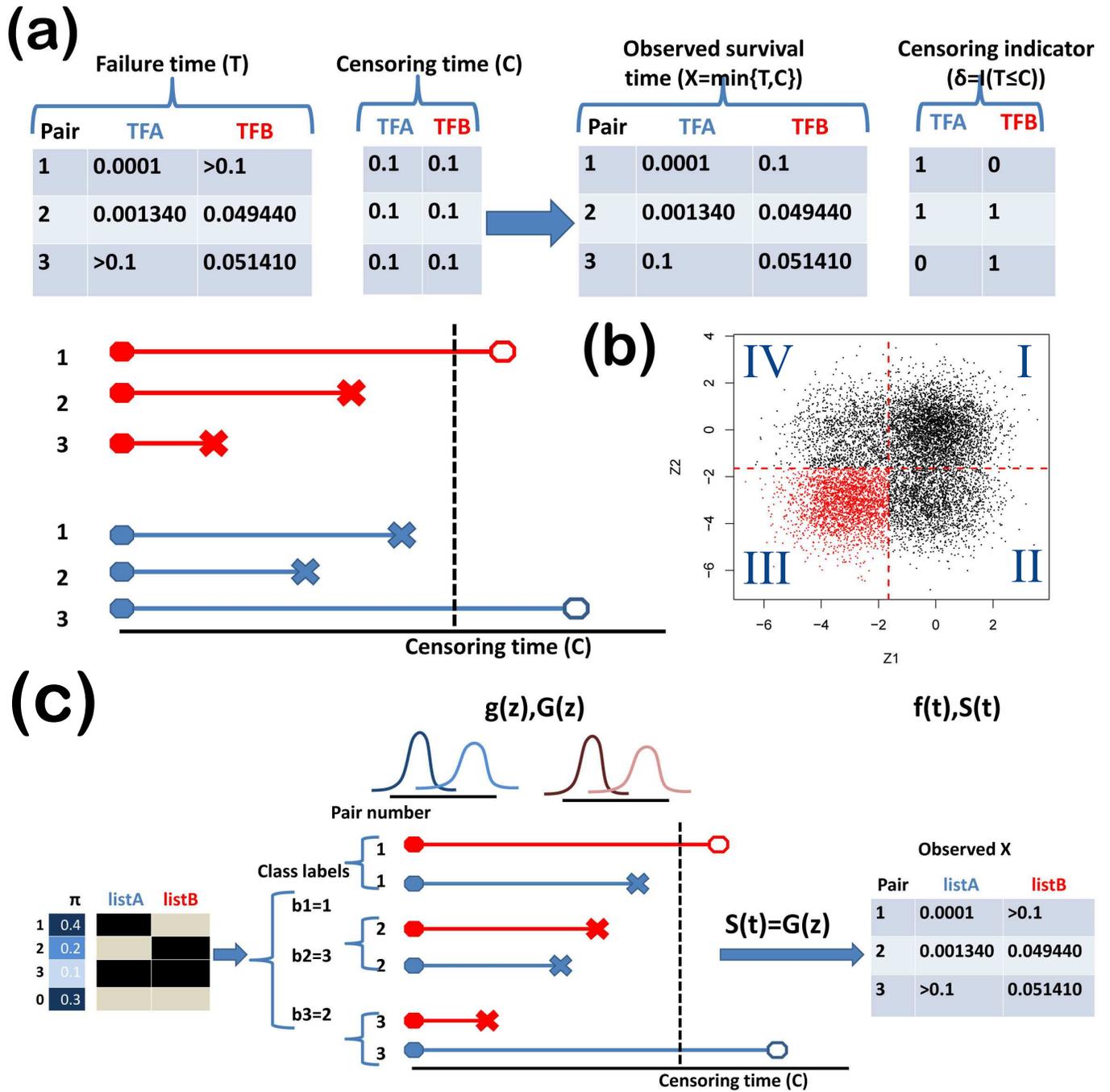}
\end{center}
\caption{The connection between the comparison of two genomic rank lists and the bivariate survival analysis. (a) An illustrative example showing how two rank lists when combined together can be represented using ``observed survival time'' and ``censoring indicators'' according to the survival terminology. (b) An illustration of how information is used in the complete case analysis by the original IDR versus the survival model in this article. (c) A cartoon illustration of the data generation process assumed by the survival copula mixture model.}
\end{figure}

The data are assumed to be generated as below (see Figure 1(c) for a cartoon illustration):
\begin{enumerate}
  \item A random indicator $b_i$ is first assigned to each locus $i$.
    \begin{itemize}
    \item  If $b_i=0$, then locus $i$ is noise in both lists.
    \item  If $b_i=1$, then locus $i$ is signal in list 1 but noise in list 2.
    \item  If $b_i=2$, then locus $i$ is signal in list 2 but noise in list 1.
    \item  If $b_i=3$, then locus $i$ is signal in both lists.
    \end{itemize}
	Thus, $b_i$ represents the co-existing pattern of signals. It is usually called ``frailty'' in survival analysis.
  The $b_i$ is assumed to be assigned according to the probability vector $\mbox{\boldmath{$\pi$}}=(\pi_0,\pi_1,\pi_2,\pi_3)$,
  where $\pi_k \equiv Pr(b_i=k)$.

\item Given $b_i$, latent variables $\tilde{z}_{i,1}$ and $\tilde{z}_{i,2}$ are generated according to $g_1(z)$ and $g_2(z)$, respectively.
\item $\tilde{z}_{i,1}$ and $\tilde{z}_{i,2}$ are truncated using $K_{1}$ and $K_{2}$ as cutoffs.
\item The truncated pseudo data $z_{i,1}$ and $z_{i,2}$ are monotone transformed to observed data $x_{i,1}$ and $x_{i,2}$ based on
    $S_j(x_{i,j})=G_j(z_{i,j})$, which yields $x_{i,j}=S_j^{-1}(G_j(z_{i,j}))$. Correspondingly, $C_{j} = S_j^{-1}(G_j(K_{j}))$.
    Also note that $x_{i,j}=min\{t_{i,j},C_{j}\}$ where and $t_{i,j} = S_j^{-1}(G_j(\tilde{z}_{i,j}))$.
\end{enumerate}

Since $T_{i,j}$ is truncated at $C_j$ and the censoring time $C_j$ is a constant, the censoring time is independent of the underlying true failure time $T_{i,j}$ and contains no information about $f_j(t)$ and $S_j(t)$. As a result, the contribution of each locus $i$ to the likelihood can be represented by one of the four formulas below:
    \begin{itemize}
    \item $b_i=0$: $h_{0i} \equiv g_{10}^{\delta_{i,1}}(z_{i,1})G_{10}^{1-\delta_{i,1}}(z_{i,1})g_{20}^{\delta_{i,2}}(z_{i,2})G_{20}^{1-\delta_{i,2}}(z_{i,2})$.
    \item $b_i=1$: $h_{1i} \equiv g_{11}^{\delta_{i,1}}(z_{i,1})G_{11}^{1-\delta_{i,1}}(z_{i,1})g_{20}^{\delta_{i,2}}(z_{i,2})G_{20}^{1-\delta_{i,2}}(z_{i,2})$.
    \item $b_i=2$: $h_{2i} \equiv g_{10}^{\delta_{i,1}}(z_{i,1})G_{10}^{1-\delta_{i,1}}(z_{i,1})g_{21}^{\delta_{i,2}}(z_{i,2})G_{21}^{1-\delta_{i,2}}(z_{i,2})$.
    \item $b_i=3$: $h_{3i} \equiv g_{11}^{\delta_{i,1}}(z_{i,1})G_{11}^{1-\delta_{i,1}}(z_{i,1})g_{21}^{\delta_{i,2}}(z_{i,2})G_{21}^{1-\delta_{i,2}}(z_{i,2})$.
    \end{itemize}
Collect the data and latent variables into three sets $\mbox{\boldmath{$Z$}}=\{z_{i,j}\}$, $\mbox{\boldmath{$\Delta$}}=\{\delta_{i,j}\}$ and $\mbox{\boldmath{$B$}}=\{b_i\}$,
and define $\mbox{\boldmath{$\theta$}}=\left\{ \mbox{\boldmath{$\pi$}},\mu_1,\mu_2,\sigma_1^2,\sigma_2^2 \right\}$.
The full likelihood can be derived as:
\begin{eqnarray}
 Pr(\mbox{\boldmath{$Z$}},\mbox{\boldmath{$\Delta$}},\mbox{\boldmath{$B$}}| \mbox{\boldmath{$\theta$}})=\prod_{i=1}^n\{\pi_0g_{10}^{\delta_{i,1}}(z_{i,1})G_{10}^{1-\delta_{i,1}}(z_{i,1})g_{20}^{\delta_{i,2}}(z_{i,2})G_{20}^{1-\delta_{i,2}}(z_{i,2})\}^{I(b_i=0)}\nonumber\\
  *\{\pi_1g_{11}^{\delta_{i,1}}(z_{i,1})G_{11}^{1-\delta_{i,1}}(z_{i,1})g_{20}^{\delta_{i,2}}(z_{i,2})G_{20}^{1-\delta_{i,2}}(z_{i,2})\}^{I(b_i=1)}\nonumber\\
  *\{\pi_2g_{10}^{\delta_{i,1}}(z_{i,1})G_{10}^{1-\delta_{i,1}}(z_{i,1})g_{21}^{\delta_{i,2}}(z_{i,2})G_{21}^{1-\delta_{i,2}}(z_{i,2})\}^{I(b_i=2)}\nonumber\\
  *\{\pi_3g_{11}^{\delta_{i,1}}(z_{i,1})G_{11}^{1-\delta_{i,1}}(z_{i,1})g_{21}^{\delta_{i,2}}(z_{i,2})G_{21}^{1-\delta_{i,2}}(z_{i,2})\}^{I(b_i=3)}\nonumber\\
  =\prod_{i=1}^n\{\pi_0h_{0i}\}^{I(b_i=0)}\{\pi_1h_{1i}\}^{I(b_i=1)}\{\pi_2h_{2i}\}^{I(b_i=2)}\{\pi_3h_{3i}\}^{I(b_i=3)}.
  \end{eqnarray}

\subsection{Model fitting}
To fit the model, we use an iterative EM algorithm similar to the one proposed by Li et al.\cite{RefLi} to estimate $\mbox{\boldmath{$\theta$}}$.
\begin{enumerate}
\item Intialize $\mbox{\boldmath{$\theta$}}$ using random values $\mbox{\boldmath{$\theta$}}_0$.
\item Use the Kaplan-Meier \cite{RefKMest} estimator to estimate the marginal survival functions for $X_{i,1}$ and $X_{i,2}$.
\item Given the initial $\mbox{\boldmath{$\theta$}}_0$, obtain pseudo-data $z_{i,1}=\hat{G}_1^{-1}(\hat{S}_1(x_{i,1}))$, $z_{i,2}=\hat{G}_2^{-1}(\hat{S}_2(x_{i,2}))$.
\item Estimate parameters $\mbox{\boldmath{$\theta$}}$ based on the pseduo-data $z_{i,1}$ and $z_{i,2}$ using an EM algorithm \cite{RefEM}.
\item Update the $\hat{G}_1$ and $\hat{G}_2$ using the newly estimated $\mbox{\boldmath{$\theta$}}$, and update the pseudo-data $z_{i,1}$ and $z_{i,2}$ using the new $\hat{G}_1$ and $\hat{G}_2$.
\item Iterate between steps 3 and 4 until the change in log-likelihood between the two nearby iterations is less than a pre-specified threshold.
\end{enumerate}
Details of the algorithm are given in the Appendix.

\subsection{Statistical Inference}
\label{s:inf}
Once the model parameters are estimated, a coexistence probability (also called probability for having reproducible signals) can be computed for each locus $i$ as:
\begin{equation}
cop(x_{i,1},x_{i,2})=Pr(K_i=3|(x_{i,1},x_{i,2}),\mbox{\boldmath{$\theta$}})=\frac{\pi_3h_{3i}}{\sum_{k=0}^3\pi_kh_{ki}}.
\end{equation}

Using these coexistence probabilities, we define two coexistence curves (COP curves) as:
\begin{equation}
COP_1(x_{i,1})=mean_{\left\{l: x_{l,1}\leq x_{i,1}\right\}}(cop(x_{l,1},x_{l,2})).
\end{equation}
\begin{equation}
COP_2(x_{i,2})=mean_{\left\{l: x_{l,2}\leq x_{i,2}\right\}}(cop(x_{l,1},x_{l,2})).
\end{equation}

Intuitively, $COP_1(x_{i,1})$ indicates that among the loci whose scores in list 1 are less than $x_{i,1}$, the proportion that are true signals in both lists. Similarly, $COP_2(x_{i,2})$ shows among the loci ranked higher than locus $i$ in list 2, the fraction that represents signals reproducible in both lists. From these two COP curves, one can see how the co-existence strengths between the two lists change from the most significant loci to the least significant ones. To facilitate the comparison with the IDR approach in \cite{RefLi}, we also define:

\begin{equation}
IDR_1(x_{i,1})=1-COP_1(x_{i,1})
\end{equation}
\begin{equation}
IDR_2(x_{i,2})=1-COP_2(x_{i,2})
\end{equation}

$IDR_1(x_{i,1})$ represents the fraction of noise or non-concordant (non-reproducible) signals among loci whose score in list 1 does not exceed $x_{i,1}$. $IDR_2(x_{i,2})$ can be interpreted similarly.
Our model and measures here allows asymmetry of the signals in the two lists. For instance, if one list is obtained from a poor quality experiment with low signal-to-noise ratio and the other list is from a high-quality experiment with high signal-to-noise ratio, the two COP curves will be different. In contrast, the original IDR approach only produces one IDR curve
to show the concordance. As a result, it cannot show the difference between two asymmetric datasets.

\section{Simulations}
In this section, we use simulations to illustrate SCOP and compare it with the Venn diagram and IDR approach.

\subsection{Characterization of degree of concordance between two rank lists}

\subsubsection{Case I}
Case I illustrates why SCOP is better at characterizing the degree of concordance between two rank lists. Consider two lists, each with 10,000 loci. Since the copula model is invariant to monotone transformation of marginal scores, we generated the simulation data by first generating latent random variables $Z_{i,1}$ and $Z_{i,2}$ and then transforming them to p-values, denoted as $T_{i,j}$, from a one-sided $z$-test for $H_0: \mu=0$ vs $H_0: \mu<0$. Specifically, $T_{i,j}=P(Z<Z_{i,j})$ where $Z$ follows the standard normal distribution $N(0,1)$. For both lists, a normal distribution $N(-5, 1)$ was used as the signal component for the latent variable $Z_{i,j}$, and $N(0,1)$ was used as the noise component. The mixture proportions of the four possible co-existence patterns were $\mbox{\boldmath{$\theta$}}=c(0.9,0,0,0.1)$. In other words, for the full lists without truncation,  only 10\% of the loci represent signals in both lists, and the other 90\% of the loci are noise. All $Z_{i,j}$ values greater than -1.65, corresponding to p-value$>0.05$ in a one-sided z-test, were truncated. The p-values $T_{i,j}$ were then generated according to the process described in Section 2.2.  Under this setting, the two lists are symmetric in terms of their signal-to-noise ratio.  To reflect the scenario in real applications, loci whose p-values were greater than $>0.05$ in both lists were excluded from the analysis. Meanwhile, all the other loci, either censored in only one or neither of the two lists, were retained.

As shown in Figure 2 (a), a total of 1,872 loci passed the p-value cutoff in either list 1 or list 2. Among them, 56.1\% (1,050) were reported in both lists. Nevertheless, the Venn diagram does not characterize the rank concordance between the two lists.

\begin{figure}
\begin{center}
\includegraphics[width=0.8\textwidth]{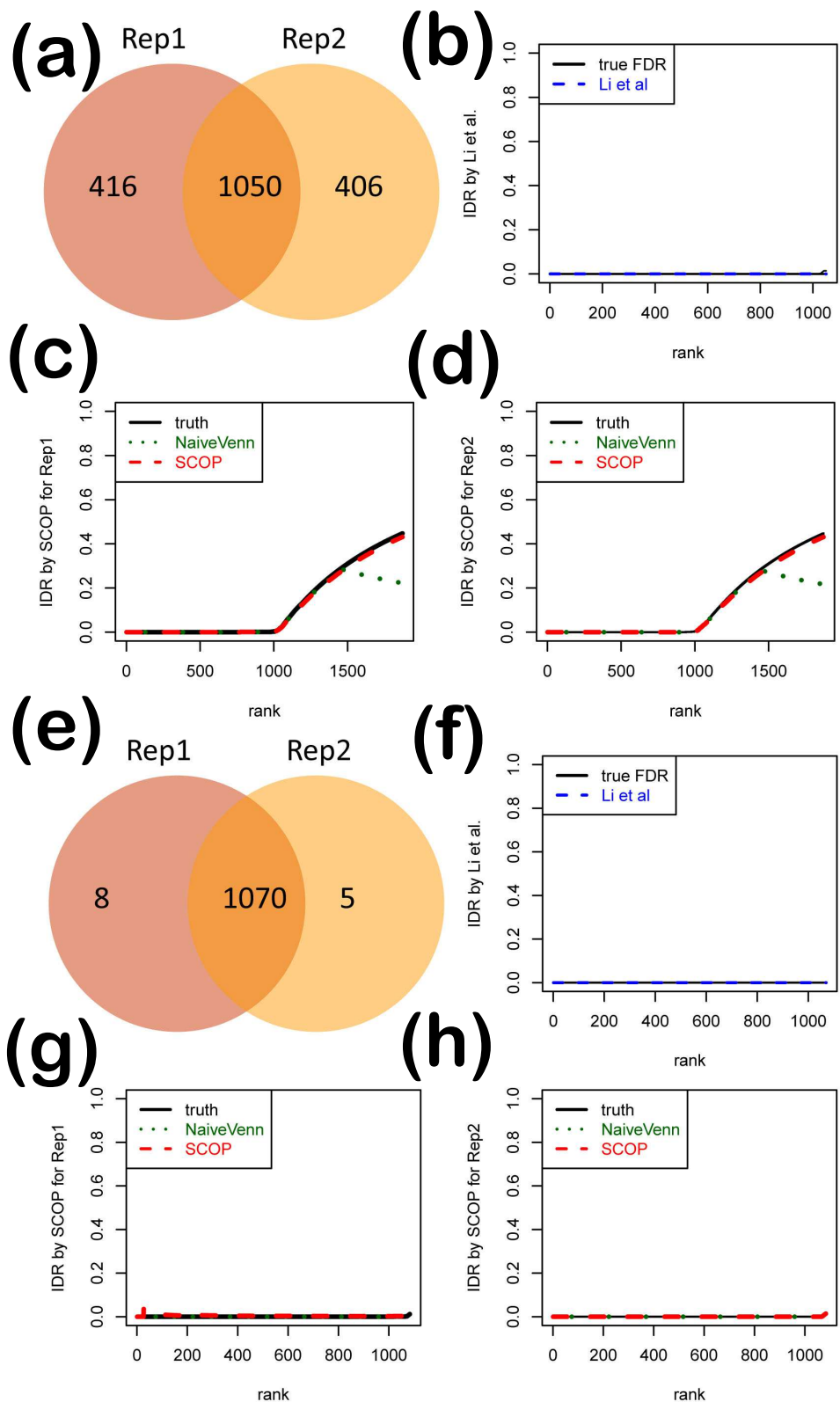}
\end{center}
\caption{(a) The Venn diagram for Case I. (b) IDR by Li et al(2011) for Case I.(c) $IDR_1$ for Case I. (d) $IDR_2$ for Case I. (e)-(f) The Venn diagram, IDR by Li et al(2011), $IDR_1$ and $IDR_2$ for Case II.}
\end{figure}

We then applied the IDR approach to the 1,050 loci reported in both lists, consistent with how the IDR analysis was performed by \cite{RefLi}.  Figure 2 (b) shows the corresponding IDR curve. Based on the curve, the IDR analysis would claim high reproducibility between the two datasets. However, this is clearly not the case, since Figure 2 (a) shows that 43.9\% of the 1,872 reported loci were not shared by the two lists. This illustrates why ignoring the list-specific loci in the IDR analysis can be misleading. The high reproducibility that the method reports only describes the degree of concordance among the loci common in both lists. It does not characterize the concordance or the reproducibility of the whole lists.
This has important implications. IDR is widely used in the ENCODE project to measure the reproducibility of replicate experiments, and to evaluate the performance of data analysis algorithms in terms of how consistent they perform when applied to replicate experiments \cite{RefEncodeGuid}. The example here shows that IDR can be very misleading if one wants to measure the global reproducibility of two replicate experiments, or to  evaluate if a data analysis algorithm is stable. This is caused by ignoring the list-specific loci, which is not allowed in the original IDR model.

Finally, we applied SCOP to the simulated data. Figures 2 (c) and (d) show the corresponding $IDR_1$ and $IDR_2$ curves, together with the underlying truth curve. The $IDR_1$ and $IDR_2$ curves (red dashed lines) match the underlying truth curves (black solid lines) very well. As a benchmark comparison, we also counted among the top ranked $k$ loci in one list how many of them were absent from the other list, and created the corresponding curves called ``NaiveVenn'' (dark green dotted lines) hereafter. From another perspective, NaiveVenn curves were constructed by fixing one circle in the Venn diagram, varying the other circle with different rank cutoffs, and counting the overlap proportions. To certain extent, ``NaiveVenn'' can be viewed as a naive estimation for $IDR_1$ and $IDR_2$. However, NaiveVenn underestimates the irreproducibility for loci occurring in only one list, whereas SCOP is able to borrowing information from all loci, complete or missing in one list, to better estimate the signal and noise proportions in the data. $IDR_1$ and $IDR_2$ curves clearly demonstrate that the fraction of concordant signal in the two lists is not high, and the irreproducible loci consist of 40\% of both observed lists. $IDR_1$ and $IDR_2$ curves also show that the signal concordance decreases as one moves from top ranked loci to lower ranked loci, a trend not directly revealed by the Venn diagram approach. With these curves, one may adjust the cutoff for calling signals based on the degree of reproducibility between two independent experiments, which is a function not usually provided by the Venn diagram approach.

\subsubsection{Case II}
In Case II, each rank list contains 1,000 loci. The mixture proportions of the co-existence patterns were $\mbox{\boldmath{$\theta$}}=c(0.1,0,0,0.9)$. For both lists, the noise and signal components for generating latent random variables $Z_{i,j}$ were again assumed to follow $N(0,1)$ and $N(-5,1)$ respectively. $Z_{i,j}$s were truncated at -1.65 as well. Among the 1,083 loci that passed the cutoff in either list, 98.8\% (1,070) were not found in both lists. Figures 2(k) and (l) show that SCOP accurately characterizes the degree of signal concordance between the two lists (compare the red and the black curves).

Comparing Figure 2(b) in Case I and Figure 2(f) in Case II, one can see that the original IDR approach would claim high reproducibility in both cases. However, Figure 2(g)(h) clearly demonstrates that these two cases are different. In Case I,  40\% of all loci are claimed as noise (Figure 2 (c)(d)) for the observed lists, whereas only 1.5\% of all loci in Case II are estimated as noise (Figure 2 (g)(h)).

In summary, the two simulations above show that the overlap revealed by Venn diagrams does not contain all information about the degree of signal concordance and how it changes when rank changes. They also show that the IDR computed using the loci present in both lists can be misleading for characterizing global concordance or reproducibility. By incorporating the censoring data into the analysis, SCOP addresses both issues and can provide a better characterization of concordance or reproducibility.

\subsection{Uncovering list-specific characteristics}
\begin{figure}
\begin{center}
\includegraphics[width=0.8\textwidth]{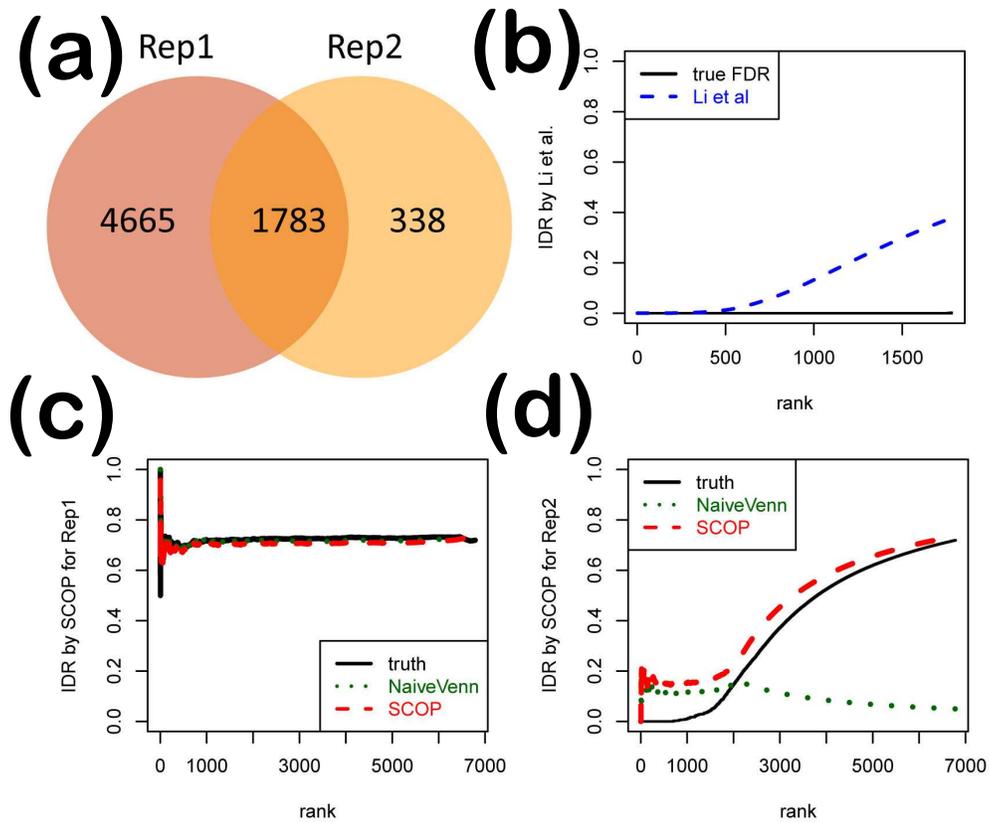}
\end{center}
\caption{(a) The Venn diagram for Case III. (b) IDR by Li et al(2011) for Case III.(c) $IDR_1$ for Case III. (d) $IDR_2$ for Case III.}
\end{figure}
Unlike the IDR approach which only produces one IDR curve, SCOP creates two curves, one for each rank list. Using these two curves, one can explore characteristics specific to each rank list. For instance,
besides measuring the overall concordance of two rank lists, IDR is also used to determine where to cut the rank lists to keep only the loci that are likely to be reproducible in independent experiments, that is, it serves a role similar to false discovery rate (FDR). In many real applications, one rank list is obtained from a high quality experiment, whereas the other list is obtained from a noisy dataset; thus, the two rank lists may be asymmetric in terms of their signal-to-noise ratio. In such a scenario, one may want to have a more detailed view of each list. Since the IDR approach produces only one IDR curve for loci shared by both lists, it does not reveal the asymmetry between the two lists. Using this IDR curve to choose cutoff forces the same cutoff to be applied to both lists. This may result in
decreased power in detecting reproducible loci. In contrast, SCOP allows one to estimate IDR separately for each list and to observe the asymmetry of data quality for the two lists.

To demonstrate, we generated scores for 10,000 loci and created two rank lists using a procedure similar to Case I in Section 3.1. For both lists, the signal and noise components were $N(-3,1)$ and $N(0,1)$ respectively. The latent variables in both lists were censored at -1.65. The mixture proportions of the four co-existence patterns were set as $\mbox{\boldmath{$\theta$}}=c(0.3,0.5,0,0.2)$. In this case, 70\% of loci in the complete list 1 are signals, of which only 20\% are also reproducible in complete list 2, corresponding to signal proportion of 96\% and 28\% in the observed lists 1 and 2. This simulation is referred to as Case III.

When the IDR approach was applied to analyze the loci present in both lists, the IDR estimates were very conservative compared to the true FDR (Figure 3(b)). This is because the asymmetry of the two lists lead to high variability, inflating the error rate estimates. In contrast, the $IDR_1$ and $IDR_2$ curves produced by SCOP accurately estimated the proportion of irreproducible signals in each list (Figure 3(c)(d)) and indicated that the two lists have asymmetric signals. Thus, a separate analysis rather than a pooled one is needed for these two lists. Moreover, Figure 3(d) once again illustrates that NaiveVenn can underestimate the irreproducibility. The reason is that it fails to distinguish between the signals and noise in the overlap part of the Venn diagram, and hence count both of them in the calculation.

\section{Real Datasets}

\subsection{Assessing reproducibility of replicate experiments}
We downloaded two replicate ChIP-seq experiments for transcription factor NF-kB in cell line Gm10847 from the ENCODE \cite{RefEncode} together with their corresponding input data (Table 1).  Peak list A was called using CisGenome \cite{RefCisgenome} by comparing sample 1 with sample 3 and 4 at cutoff of FDR=0.01; similarly, peak list B was called by comparing sample 2 with sample 3 and 4. For each peak, we extracted the 150bp window centered at the peak summit to ensure the same length in peaks. We then compared the two peak lists. The Venn diagram in Figure 4(a) shows that the majority of the loci in these two peak lists were different. Only 39.0\% of loci in list A and 20.5\% of loci in list B were found in the other list. Nevertheless, the IDR analysis of the shared loci gives a low IDR estimate of 0.015, misleadingly suggesting high reproducibility between the two replicates (Figure 4(b)). In contrast, SCOP was able to show that the two replicate experiments have low reproducibility and high IDRs (Figure 4(c)(d)).

\begin{table}
\scalebox{0.9}{
\begin{tabular}{cllll}
\hline Id& Section & List name & File name & Experiment type\\
\hline
1&$4.1$ & RepA &wgEncodeSydhTfbsGm10847NfkbTnfaIggrabAlnRep2.bam & NF-kB \\
2&&RepB & wgEncodeSydhTfbsGm10847NfkbTnfaIggrabAlnRep4.bam & NF-kB \\
3&& &wgEncodeSydhTfbsGm10847InputIggmusAlnRep1.bam & Input \\
4&& & wgEncodeSydhTfbsGm10847InputIggmusAlnRep2.bam & Input \\
5&$4.2$ & FoxA1 &wgEncodeHaibTfbsHepg2Foxa1sc6553V0416101AlnRep1.bam & FoxA1 \\
6&& FoxA1 &wgEncodeHaibTfbsHepg2Foxa1sc6553V0416101AlnRep2.bam & FoxA1 \\
7&& FoxA2 &wgEncodeHaibTfbsHepg2Foxa2sc6554V0416101AlnRep1.bam& FoxA2 \\
8&& FoxA2 &wgEncodeHaibTfbsHepg2Foxa2sc6554V0416101AlnRep2.bam& FoxA2 \\
9&& &wgEncodeHaibTfbsHepg2RxlchV0416101AlnRep1.bam& Input \\
10&& &wgEncodeHaibTfbsHepg2RxlchV0416101AlnRep2.bam& Input \\
\hline
\end{tabular}}
\caption{Data description for real ENCODE ChIP-seq datasets.}
\end{table}

\begin{figure}
\begin{center}
\includegraphics[width=0.8\textwidth]{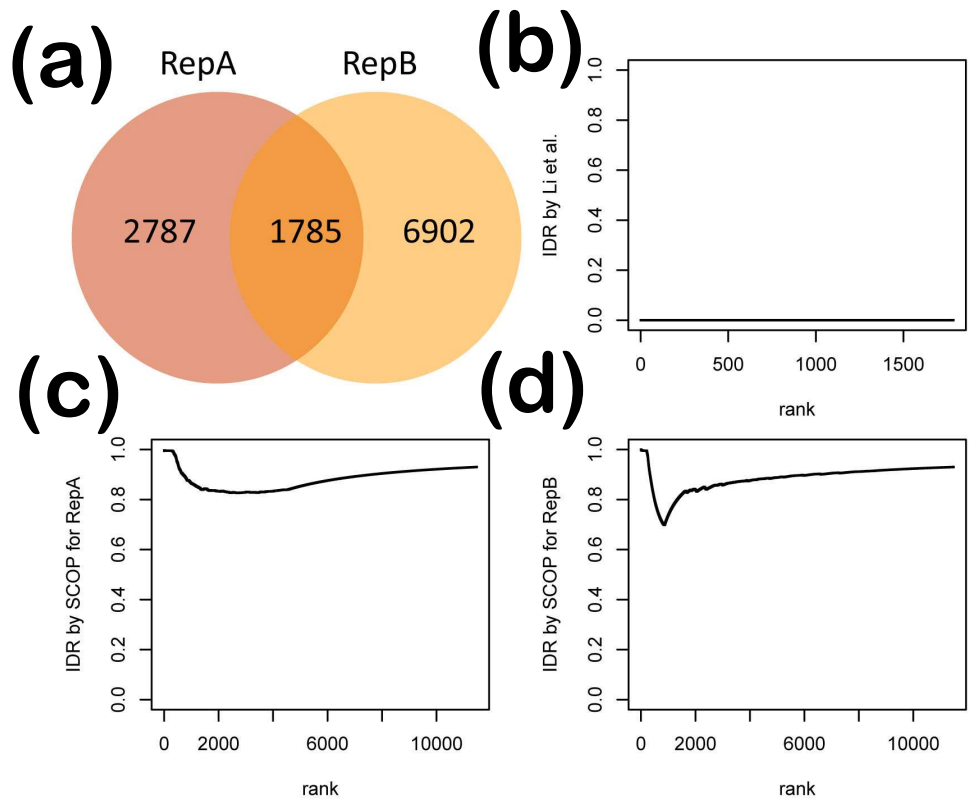}
\end{center}
\caption{Different measures for two peak lists constructed from NF-kB ChIP-seq datasets in cell line Gm10847 from the ENCODE project. (a) The Venn diagram. (b) IDR by Li et al(2011).(c) $IDR_1$ for Replicate A. (d) $IDR_2$ for Replicate B.}
\end{figure}

\subsection{Characterizing co-binding of two transcription factors (TF)}
FoxA transcription factors are a key family of TFs that regulate gene activities in liver cancer. Biologists are interested in how members in this TF family interact with each other and whether different members bind to the same genomic loci in liver cancer cells. The ENCODE project has generated ChIP-seq data for both FoxA1 and FoxA2 in a liver cancer cell line Hepg2. These data can be used to answer the questions raised in Section 1. Using CisGenome \cite{RefCisgenome}, we called 65,535 binding peaks for FoxA1 (comparing sample 5 and 6 with sample 9 and 10) and 48,503 peaks for FoxA2 (comparing sample 7 and 8 with sample 9 and 10), respectively at the FDR=0.01 cutoff.

\begin{figure}
\begin{center}
\includegraphics[width=0.8\textwidth]{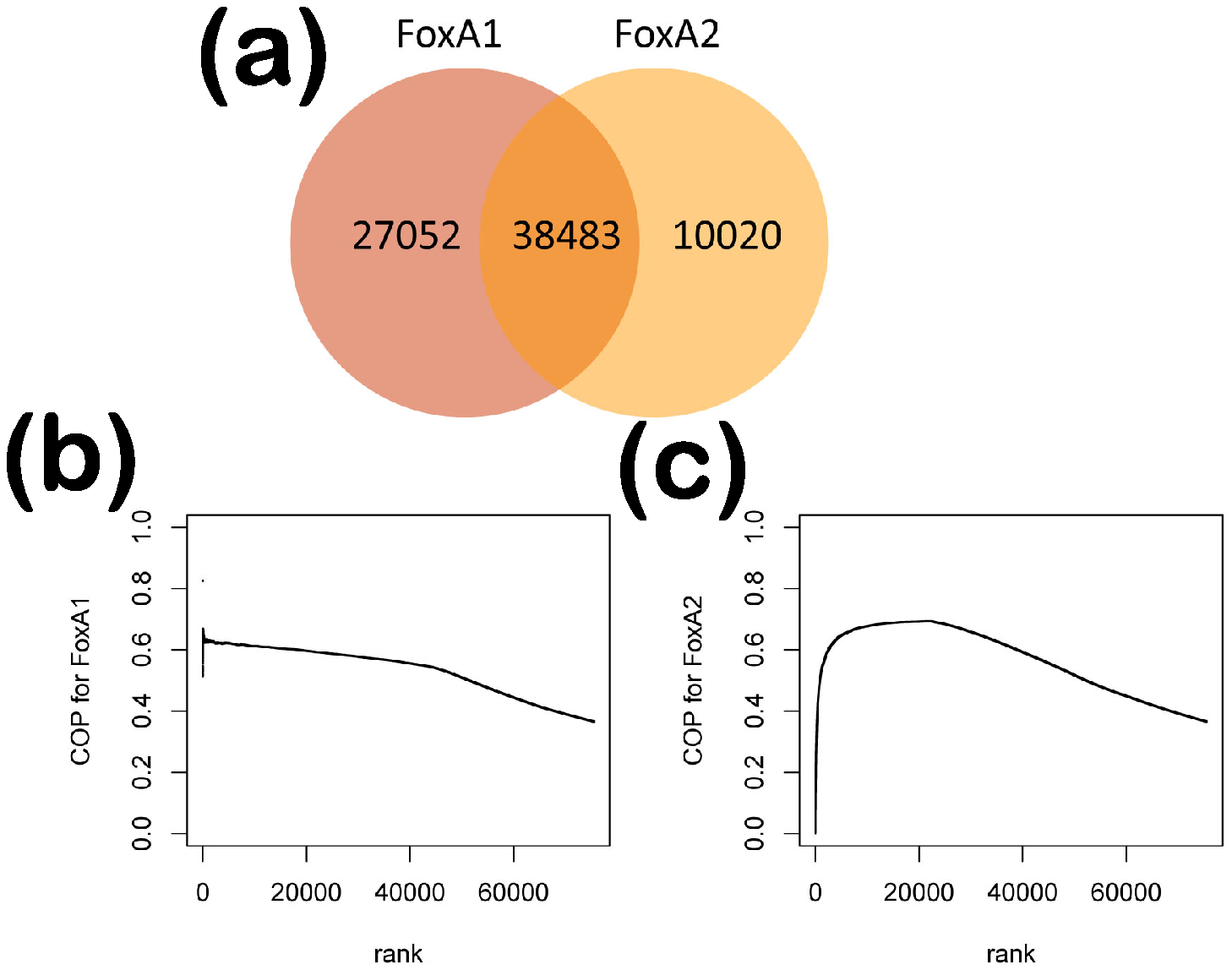}
\end{center}
\caption{COP curves for FoxA1 and FoxA2 peak lists constructed from ChIP-seq datasets in Hepg2 cell line from the ENCODE project.}
\end{figure}

Finally, we applied SCOP to characterize the concordance between the two lists.
Figure 5(a) shows that among the top ranked FoxA1 peak regions, about 60\% are also bound by FoxA2. As one moves to the lower ranked FoxA1 peaks, a lower percentage are simultaneously bound by FoxA2. Thus, robust FoxA1 binding seems to require FoxA2 binding at the same location. In contrast, the very top ranked FoxA2 peaks are more likely to be FoxA2 specific and less likely to be shared for FoxA1 binding. The middle ranked FoxA2 peaks are more often bound by FoxA1. The co-binding proportion drops again for FoxA2 peaks with low ranking which are increasingly more likely to be noise. This suggests that FoxA2 may play its regulatory role in a different mode compared to FoxA1. This information is not immediately revealed by the Venn diagram and IDR approach.

\section{Discussion}
\label{s:discuss}
In summary, SCOP offers a new solution for comparing two rank lists. SCOP takes into account both the overall proportion of overlap shared by the two lists and the consistency of ranks along them. This overcomes the shortcomings of the Venn diagram and the IDR approach, and allows better characterizing of the concordance and global reproducibility between two datasets. Our simulation studies show that drawing conclusions on concordance from Venn diagrams may not reveal all the information in the data. The same degree of overlap may correspond to different signal-to-noise ratio. IDR, on the other hand, is limited in terms of characterizing the global reproducibility between two datasets since it focuses on analyzing loci shared by both lists. In light of these results, the SCOP curves should provide a better solution to assessing data quality (e.g., reproducibility between replicate ChIP-seq samples) and computational algorithms (e.g., evaluate consistency of the results when a method is applied to two replicate experiments) in projects such as ENCODE.

Our current model considers the problem of comparing two rank lists. An interesting future research topic is how to extend it to comparing multiple rank lists. Currently, one can apply SCOP to compare each pair of lists. However, this pairwise comparison approach does not directly reveal higher order relationships. For instance, with three datasets, one can also ask how many loci are shared by all three lists in addition to asking how many loci are shared by each pair of lists. For $D$ rank lists, there are $2^D$ combinatorial  signal coexistence patterns. As $D$ increases, the complexity of the problem increases exponentially. Efficient ways to perform the comparison and summarize results, similar to those in \cite{RefiASeq}, need to be developed in order to solve this problem.

Currently, an R package for SCOP is available upon request. The package will soon be submitted to Bioconductor.

\setcounter{section}{0}
\setcounter{equation}{0}
\setcounter{table}{0}
\renewcommand{\theequation}{A.\arabic{equation}}
\renewcommand{\thetable}{A.\arabic{table}}
\renewcommand{\thefigure}{A.\arabic{figure}}

\section*{Appendix}

\subsection*{Iterative algorithm for model fitting}\label{app}

Here we present the details of the iterative algorithms used to estimate $\mbox{\boldmath{$\theta$}}=(\mbox{\boldmath{$\pi$}},\mu_1,\mu_2,\sigma_1^2,\sigma_2^2)$.

\begin{enumerate}
\item Initialize parameters $\mbox{\boldmath{$\theta$}}=\mbox{\boldmath{$\theta_0$}}$.

\item Estimate the survival function $S_j(x_{i,j})$ using the Kaplan-Meier estimator.

\item Compute the pseudo-data $\hat{z}_{i,j}=G^{-1}_j(\hat{S}_{j}(x_{i,j})|\mbox{\boldmath{$\theta$}})$. Since $G^{-1}_j$ does not have a closed form, $G_j$ is first computed on a grid of 5,000 points over the range $[min(-5,\mu_j-5*\sigma_j),max(5,\mu_j+5*\sigma_j)]$. $\hat{z}_{i,j}$ is then obtained through linear interpolation on the grid.

\item Run EM algorithm to search for $\hat{\mbox{\boldmath{$\theta$}}}$ that maximizes the log-likelihood of pseudo data
 $Pr(\hat{\mbox{\boldmath{$Z$}}},\mbox{\boldmath{$\Delta$}}| \mbox{\boldmath{$\theta$}})=\sum_{\mbox{\boldmath{$B$}}}Pr(\hat{\mbox{\boldmath{$Z$}}},\mbox{\boldmath{$\Delta$}},\mbox{\boldmath{$B$}}| \mbox{\boldmath{$\theta$}})$. The resulting $\hat{\mbox{\boldmath{$\theta$}}}$ is denoted as $\mbox{\boldmath{$\theta^t$}}$.

5. Iterate between steps 3 and 4 until the change in log-likelihood between the two nearby iterations is less than a pre-specified threshold.
\end{enumerate}

Below are details of the EM algorithm in step 4. In the E-step, one evaluates the Q-function
\begin{equation}
Q(\mbox{\boldmath{$\theta$}}|\mbox{\boldmath{$\theta^{old}$}})=Q(\mbox{\boldmath{$\pi$}},\mu_1,\mu_2,\sigma_1^2,\sigma_2^2| \mbox{\boldmath{$\pi^{old}$}},\mu_1^{old},\mu_2^{old},\sigma_1^{2old},\sigma_2^{2old})=E_{old}(\ln Pr(\hat{\mbox{\boldmath{$Z$}}},\mbox{\boldmath{$\Delta$}},\mbox{\boldmath{$B$}}| \mbox{\boldmath{$\theta^{old}$}}))
\end{equation}
Here the expectation is taken with respect to probability distribution $Pr(\mbox{\boldmath{$B$}}| \hat{\mbox{\boldmath{$Z$}}},\mbox{\boldmath{$\Delta$}},\mbox{\boldmath{$\theta^{old}$}})$.

\begin{eqnarray}
    \ln Pr(\hat{\mbox{\boldmath{$Z$}}},\mbox{\boldmath{$\Delta$}},\mbox{\boldmath{$B$}}| \mbox{\boldmath{$\theta$}})=\sum_{i=1}^nI(b_i=0)*\{\ln\pi_0+\delta_{i,1}\ln g_{10}(\hat{z}_{i,1})+(1-\delta_{i,1})\ln G_{10}(\hat{z}_{i,1})\nonumber\\
    +\delta_{i,2}\ln g_{20}(\hat{z}_{i,2})+(1-\delta_{i,2})\ln G_{20}(\hat{z}_{i,2})\}\nonumber\\
  +(b_i=1)*\{\ln\pi_1+\delta_{i,1}\ln g_{11}(\hat{z}_{i,1})+(1-\delta_{i,1})\ln G_{11}(\hat{z}_{i,1})\nonumber\\
    +\delta_{i,2}\ln g_{20}(\hat{z}_{i,2})+(1-\delta_{i,2})\ln G_{20}(\hat{z}_{i,2})\}\nonumber\\
  +I(b_i=2)*\{\ln\pi_2+\delta_{i,1}\ln g_{10}(\hat{z}_{i,1})+(1-\delta_{i,1})\ln G_{10}(\hat{z}_{i,1})\nonumber\\
    +\delta_{i,2}\ln g_{21}(\hat{z}_{i,2})+(1-\delta_{i,2})\ln G_{21}(\hat{z}_{i,2})\}\nonumber\\
  +I(b_i=3)*\{\ln\pi_3+\delta_{i,1}\ln g_{11}(\hat{z}_{i,1})+(1-\delta_{i,1})\ln G_{11}(\hat{z}_{i,1})\nonumber\\
    +\delta_{i,2}\ln g_{21}(\hat{z}_{i,2})+(1-\delta_{i,2})\ln G_{21}(\hat{z}_{i,2})\}.
  \end{eqnarray}

Therefore,

\begin{eqnarray}
    Q(\mbox{\boldmath{$\theta$}}|\mbox{\boldmath{$\theta^{old}$}})&=&\sum_{i=1}^n\sum_{k=0}^3P_{old}(b_i=k)\ln\pi_k\nonumber\\
   &+&\sum_{i=1}^n\{(Pr_{old}(b_i=1)+Pr_{old}(b_i=3))(\delta_{i,1}\ln g_{11}(\hat{z}_{i,1})+(1-\delta_{i,1})\ln G_{11}(\hat{z}_{i,1}))\nonumber\\
   &&+(Pr_{old}(b_i=0)+Pr_{old}(b_i=2))(\delta_{i,1}\ln g_{10}(\hat{z}_{i,1})+(1-\delta_{i,1})\ln G_{10}(\hat{z}_{i,1}))\}\nonumber\\
   &&+(Pr_{old}(b_i=2)+Pr_{old}(b_i=3))(\delta_{i,2}\ln g_{21}(\hat{z}_{i,2})+(1-\delta_{i,2})\ln G_{21}(\hat{z}_{i,2}))\nonumber\\
   &&+(Pr_{old}(b_i=0)+Pr_{old}(b_i=1))(\delta_{i,2}\ln g_{20}(\hat{z}_{i,2})+(1-\delta_{i,2})\ln G_{20}(\hat{z}_{i,2}))\nonumber\\.
  \end{eqnarray}

In the M-step, one finds $\mbox{\boldmath{$\theta$}}$ that maximize the Q-function
$Q(\mbox{\boldmath{$\theta$}}|\mbox{\boldmath{$\theta^{old}$}})$. Denote them by
$\mbox{\boldmath{$\hat{\theta}^{new}$}}=(\mbox{\boldmath{$\pi^{new}$}},\mu_1^{new},\mu_2^{new},\sigma_1^{2new},\sigma_2^{2new})$.
Solving
\begin{equation}
 \frac{\partial Q(\mbox{\boldmath{$\theta$}}|\mbox{\boldmath{$\theta^{old}$}})}{\partial \pi_k}=0
\end{equation}
We have:
\begin{equation}
\hat{\pi}_k^{new}=\frac{\sum_{i=1}^nPr_{old}(b_i=k)}{n}
\end{equation}

Recall:
\begin{eqnarray}
Pr_{old}(b_i=k)&=&Pr(b_i=k|\hat{z}_{i,1},\hat{z}_{i,2},\delta_{i,1},\delta_{i,2},\mbox{\boldmath{$\hat{\theta}^{old}$}})\\
&=&\frac{Pr(b_i=k,\hat{z}_{i,1},\hat{z}_{i,2}|\delta_{i,1},\delta_{i,2},\mbox{\boldmath{$\hat{\theta}^{old}$}})}{Pr(\hat{z}_{i,1},\hat{z}_{i,2}|\delta_{i,1},\delta_{i,2},\mbox{\boldmath{$\hat{\theta}^{old}$}})}\nonumber\\
\nonumber
\end{eqnarray}
and
\begin{equation}
Pr(b_i=0,\hat{z}_{i,1},\hat{z}_{i,2}|\delta_{i,1},\delta_{i,2},\mbox{\boldmath{$\hat{\theta}$}})=\pi_0g_{10}^{\delta_{i,1}}(\hat{z}_{i,1})G_{10}^{1-\delta_{i,1}}(\hat{z}_{i,1})g_{20}^{\delta_{i,2}}(\hat{z}_{i,2})G_{20}^{1-\delta_{i,2}}(\hat{z}_{i,2})
\end{equation}

\begin{equation}
Pr(b_i=1,\hat{z}_{i,1},\hat{z}_{i,2}|\delta_{i,1},\delta_{i,2},\mbox{\boldmath{$\hat{\theta}$}})=\pi_1g_{11}^{\delta_{i,1}}(\hat{z}_{i,1})G_{11}^{1-\delta_{i,1}}(\hat{z}_{i,1})g_{20}^{\delta_{i,2}}(\hat{z}_{i,2})G_{20}^{1-\delta_{i,2}}(\hat{z}_{i,2})
\end{equation}

\begin{equation}
Pr(b_i=2,\hat{z}_{i,1},\hat{z}_{i,2}|\delta_{i,1},\delta_{i,2},\mbox{\boldmath{$\hat{\theta}$}})=\pi_2g_{10}^{\delta_{i,1}}(\hat{z}_{i,1})G_{10}^{1-\delta_{i,1}}(\hat{z}_{i,1})g_{21}^{\delta_{i,2}}(\hat{z}_{i,2})G_{21}^{1-\delta_{i,2}}(\hat{z}_{i,2})
\end{equation}

\begin{equation}
Pr(b_i=3,\hat{z}_{i,1},\hat{z}_{i,2}|\delta_{i,1},\delta_{i,2},\mbox{\boldmath{$\hat{\theta}$}})=\pi_3g_{11}^{\delta_{i,1}}(\hat{z}_{i,1})G_{11}^{1-\delta_{i,1}}(\hat{z}_{i,1})g_{21}^{\delta_{i,2}}(\hat{z}_{i,2})G_{21}^{1-\delta_{i,2}}(\hat{z}_{i,2})
\end{equation}

$Pr_{old}(b_i=k)$ can be computed by replacing $\mbox{\boldmath{$\hat{\theta}$}}$ with $\mbox{\boldmath{$\hat{\theta}^{old}$}}$ accordingly.

Only $\sum_{i=1}^n(Pr_{old}(b_i=1)+Pr_{old}(b_i=3))(\delta_{i,1}\ln g_{11}(\hat{z}_{i,1})+(1-\delta_{i,1})\ln G_{11}(\hat{z}_{i,1}))$ in Equation A.3 involves $(\mu_1,\sigma_1)$. Because $G_{11}(\hat{z}_{i,1})$, the tail probability of a normal distribution, has no close form, we use the R function $optim$ with the ``L-BFGS-B'' option to obtain the values that maximize $\sum_{i=1}^n(Pr_{old}(b_i=1)+Pr_{old}(b_i=3))(\delta_{i,1}\ln g_{11}(\hat{z}_{i,1})+(1-\delta_{i,1})\ln G_{11}(\hat{z}_{i,1}))$. $(\mu_2,\sigma_2)$ are searched in a similar fashion to maximize $\sum_{i=1}^n(Pr_{old}(b_i=2)+Pr_{old}(b_i=3))(\delta_{i,2}\ln g_{21}(\hat{z}_{i,2})+(1-\delta_{i,2})\ln G_{21}(\hat{z}_{i,2}))$.

\section*{Acknowledgements}

The authors thank Professor Mei-Cheng Wang and members from the Hopkins SLAM and Genomics working group for their helpful discussions and suggestions.

\end{document}